# Astro2020 Science White Paper

# Imprint of Drivers of Galaxy Formation in the Circumgalactic Medium

**Thematic Areas:**

- ☐ Planetary Systems
- ☐ Formation and evolution of compact objects
- ☐ Stars and Stellar Evolution
- ☑ Galaxy Evolution
- ☐ Star and Planet formation
- ☐ Cosmology and Fundamental Physics
- ☐ Resolved stellar populations and their environments
- ☐ Multi-Messenger Astronomy and Astrophysics

**Authors:** Benjamin D. Oppenheimer (*University of Colorado, Boulder*),
Juna Kollmeier (*Carnegie Observatories*), Andrey Kravtsov (*University of Chicago*),
Joel Bregman (*University of Michigan*), Daniel Anglés-Alcázar (*Flatiron Institute*),
Robert Crain (*Liverpool John Moores University*), Romeel Davé (*Royal Observatory of Edinburgh*),
Lars Hernquist (*Harvard-Smithsonian*), Cameron Hummels (*Caltech*),
Joop Schaye (*Leiden University*), Grant Tremblay (*Harvard-Smithsonian*),
G. Mark Voit (*Michigan State University*), Rainer Weinberger (*Harvard-Smithsonian*),
Jessica Werk (*University of Washington*), Nastasha Wijers (*Leiden University*),
John A. ZuHone (*Harvard Smithsonian*), Akos Bogdan (*Harvard-Smithsonian*),
Ralph Kraft (*Harvard-Smithsonian*), Alexey Vikhlinin (*Harvard-Smithsonian*)

**Contact:** Benjamin D. Oppenheimer
Institution: University of Colorado, Boulder
Email: benjamin.oppenheimer@colorado.edu
Phone: 303-492-5592

*Executive Summary.* The majority of baryons reside beyond the optical extent of a galaxy in the circumgalactic and intergalactic media (CGM/IGM). Gaseous halos are inextricably linked to the appearance of their host galaxies through a complex story of accretion, feedback, and continual recycling. The energetic processes, which define the state of gas in the CGM, are the same ones that 1) regulate stellar growth so that it is not over-efficient, and 2) create the diversity of today's galaxy colors, star formation rates, and morphologies spanning Hubble's Tuning Fork Diagram. They work in concert to set the speed of growth on the star-forming Main Sequence, transform a galaxy across the Green Valley, and maintain a galaxy's quenched appearance on the Red Sequence.

We are in an era when UV absorption studies are dramatically increasing our knowledge of the CGM gas. These revolutionary observations are shifting focus to the CGM as one of the most crucial probes of galaxy evolution. However, the majority of baryons in halos more massive than $\sim 10^{12} \, M_\odot$ along with their physics and dynamics remain invisible because that gas is heated above the UV ionization states. We argue that information on many of the essential drivers of galaxy evolution is primarily contained in this "missing" hot gas phase.

Completing the picture of galaxy formation requires uncovering the physical mechanisms behind stellar and super-massive black hole (SMBH) feedback driving mass, metals, and energy into the CGM. By opening galactic hot halos to new wavebands, we not only obtain fossil imprints of > 13 Gyrs of evolution, but we can observe on-going hot-mode accretion, the deposition of superwind outflows into the CGM, and the re-arrangement of baryons by SMBH feedback. A description of the flows of mass, metals, and energy will only be complete by observing the thermodynamic states, chemical compositions, structure, and dynamics of $T \geq 10^6$ K halos. These measurements are uniquely possible with a next-generation X-ray observatory if it provides the sensitivity needed to detect faint CGM emission, spectroscopic power to detect spectral lines and measure gas motions, and high spatial resolution to resolve structures.

*Introduction: The Gaseous Halos around Today's Galaxies* An intimate connection between properties of the stellar component in galaxies and their extended gas halos has long been predicted by analytic theories of galaxy formation. A transition of galaxy properties at the $\sim 10^{12} \, M_\odot$ dark matter (DM) halo mass scale can be related to the maximum in the baryonic cooling curve at the corresponding virial temperatures [1, 2]. The formation of a hot ($T > 10^6$ K), ambient gaseous halo should accompany a decline in the efficiency of accretion onto a galaxy, and of star formation within it [3–5]. These analytic models also revealed one of the central questions, the so-called "over-cooling" problem: cooling and condensation of the halo gas averaged over cosmic time is expected to proceed at much higher rates than the observed star formation [6]. Likely solutions to the over-cooling problems leave an imprint in the circumgalactic medium (CGM). Galactic-scale feedback processes driven by stellar winds, radiation pressure, and explosive outflows [e.g. 7–9], and by radiation, jets, and thermal bubbles from supermassive black holes (SMBH) [e.g. 10–12] eject baryons from the galactic sites of star formation into the CGM.

UV-based CGM observations afforded by the *Hubble*'s Cosmic Origins Spectrograph (COS) has redefined our understanding of galactic halos. UV absorption lines from hydrogen and metals are ubiquitous in quasar spectra intersecting the CGM. COS observed multi-phase gaseous content of galactic halos, revealing a rich set of dynamical processes — gas accretion, outflows, and recycling that correlate with galaxy type [13, and references therein].

State-of-the-art cosmological hydrodynamic simulations have applied increasingly sophisticated feedback schemes tied to star formation and SMBH accretion [e.g. 14–16]. They reproduce detailed properties of galaxies and provide very specific predictions for the physical properties of the CGM. Basic theory — $T_{\rm vir} \approx 10^6 \times (M_{\rm halo}/10^{12} M_\odot)^{2/3}$ K — and multiple simulation suites, including EAGLE

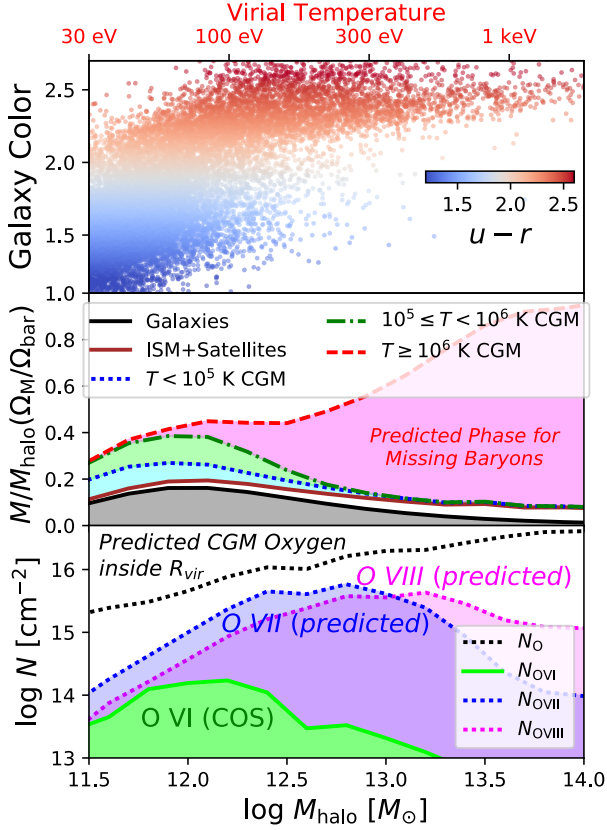

FIG. 1 Current observations and theory predictions indicate a dramatic transition in the properties of galaxies at a mass scale of ≈ the Milky Way. **Top:** Low-$z$ galaxies show the greatest variety of colors, SFRs, and morphologies in the mass range $(1 - 3) \times 10^{12} M_\odot$ [22]. **Middle:** Abundance matching analyses [23] show that star formation efficiency peaks around $10^{12} M_\odot$. A decline at higher masses is thought to be associated with the growing importance of hot baryons (red) over the cool [$T < 10^5$ K, 21] and warm CGM [$10^5 \leq T < 10^6$ K, 24] CGM phases. The white area above the red dashed line are the expected baryons ejected beyond $R_{\rm vir}$. **Bottom:** The average column density of oxygen in different ionization states within $R_{\rm vir}$. While O VI has been observed around normal galaxies with COS [24], much larger reservoirs of O VII and O VIII, accessible only via an X-ray telescope, dominate the predicted CGM oxygen budget [25–27]. *Shaded regions in these panels are theoretical estimates, currently highly unconstrained.*

[17], Illustris-TNG [18], and FIRE-2 [19] all predict that the majority of CGM gas and metals in the hot phase in halos at or above the Milky Way (MW) mass scale, $M_{\rm halo} \approx 10^{12} M_\odot$ (Fig. 1). At least some simulation suites [e.g., EAGLE, see 20] simultaneously reproduce the UV absorption line statistics for the cool/warm ($T \leq 10^{5.5}$ K) phase of the CGM [e.g. 21]. Therefore, we expect that the "COS era" is just the dawn of our understanding of the baryon physics regulating, transforming, and quenching galaxies. Despite the rich array of ions detected by UV absorption, the majority of CGM baryons and metals remain undetected in the hot ($T \geq 10^6$ K) phase. Their physics, dynamics, and energetics contain essential clues on how galaxies assemble, evolve, and transform.

*Major Questions: How Galaxies Evolve, Transform, and Quench?* A full description of formation and evolution of galaxies depends on revealing the high-energy processes that operate on vastly different scales. Critical gaps of knowledge remain unconstrained without observing the accretion physics onto $10^{5-8} M_\odot$ SMBHs, the energetic yet diffuse stellar and SMBH-driven superwinds emanating from galactic discs, and the complex tapestry of mass, metal, and energy flows exchanging phases in the CGM. The fundamental relations between galaxies, their SMBHs, and *all* phases of the CGM are currently in the realm of competing and unconstrained theoretical models. It is essential that new observational capabilities are developed to enable their direct study.

The current generation of simulations all agree that a confluence of high-energy processes operate to evolve and transform a galaxy's optically observed properties: color, morphology, SFR. A blue cloud galaxy regulates its star formation through a balance of accretion from the CGM (both cold, $T \sim 10^4$ K, and hot, $T > 10^5$ K), and star formation-driven superwind outflows. The formation of a virialized hot halo at $M_{\rm vir} \sim 10^{12} M_\odot$ curtails this cycle as the gas cooling becomes inefficient. At this stage, the SMBH is predicted to grow much faster than the galaxy due to the decline of bursty star formation-driven winds [28] and/or the collection of nuclear hot gas [29] leading to AGN feedback that drives jets, shocks, and bubbles 10's to 100's of kpc into the CGM [e.g., 30]. This amount of energy and momentum imparted to MW-mass gaseous halos can 1) shred the cool/warm CGM accretion supply, 2) uplift and eject baryons from the CGM [31], and 3) secularly transform a galaxy across the "green valley" [32]. The re-arranged quasi-stable, high entropy halo, from which accretion becomes inefficient, leads to the transition of the central galaxy to the red sequence, where maintenance-mode AGN feedback may be required to maintain the galaxy's quenched appearance. The main goal of future observational programs should be to show how these processes work in concert to set the speed of growth on the star-forming Main Sequence, transform a galaxy across the Green Valley, and maintain a galaxy's quenched appearance on the Red Sequence.

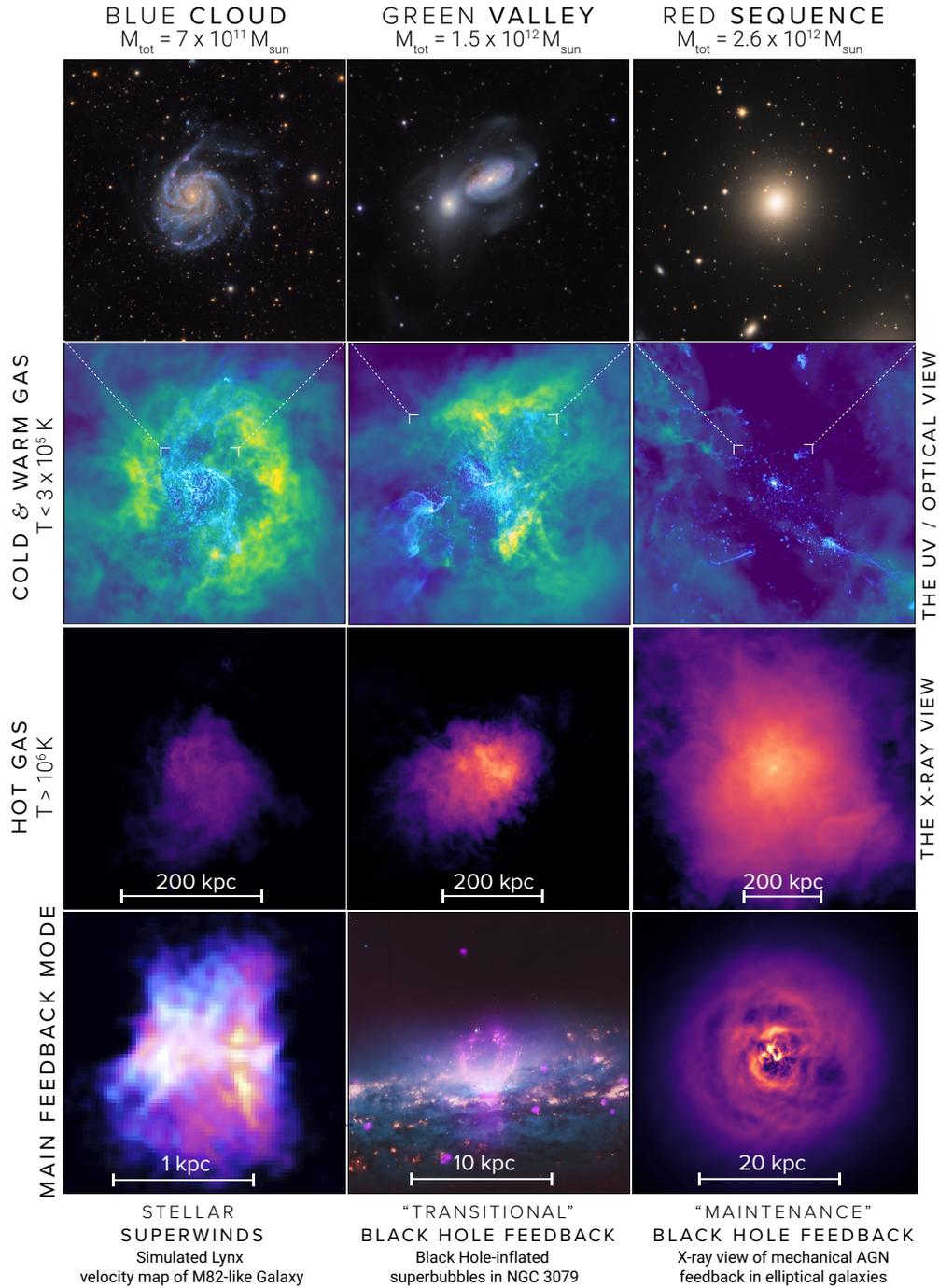

FIG. 2 A drastic transformation of galaxies from star-forming blue cloud to green valley galaxy and further to red sequence happens in a relatively narrow range of masses around $M_{\text{tot}} = 1 \times 10^{12} M_\odot$. The diverse optical appearances (top panels) are intimately tied to their circumgalactic gas reservoirs. Numerical simulations predict rich structure of all CGM phases (2nd and 3rd panel rows show outputs from three EAGLE Super-HiRes zooms). This structure reflects crucial "invisible drivers" of galaxy formation: accretion, feedback, and recycling flows of gas. The CGM in its full glory remains largely unobserved. The structure of cold ($T < 10^{4.5}$ K, white-blue) and warm ($T \sim 10^{5.5}$ K, green-yellow) gas is probed via UV absorption in individual sight lines; hot phase is routinely observed in the X-rays only in more massive ellipticals. As the galaxy mass increases, we expect a dramatic transformation of the CGM from warm-cold dominated for $M < 10^{12} M_\odot$ to hot-dominated for $M \gtrsim 2 \times 10^{12} M_\odot$. Energy feedback, which plays a major role in this transformation, changes predominant types around the $M_{\text{tot}} = 1 \times 10^{12} M_\odot$ mass scale (bottom panels). In all cases, the main signatures of on-going feedback are imprinted in the inner structure of the hot halo. Their observations require a combination of high spatial and high spectral resolution uniquely provided by *Lynx*.

***Observing Tool of the Future: hidden reservoirs of mass, metals, and energy.*** The same energetic processes that 1) regulate galaxy assembly so that it is not over-efficient, and 2) create the diversity of today's galaxy colors, SFR's & morphologies spanning Hubble's Tuning Fork Diagram, also define the state of gas in the CGM by heating and ejecting mass and metals into it. This makes the CGM properties exceptionally powerful in revealing these "invisible drivers" of galaxy formation.

As we noted above, the majority of baryons and metals remain undetected in $\geq 10^{12} M_\odot$ halos [33], most likely because they are in the hot phase. How much could a future observatory find? The maximum fraction of baryons locked in stars is ~ 20%, and is found in MW-mass halos (Fig. 1, middle panel). Budgeting the baryons out to the virial radius of MW-mass halos ($R_{\rm vir}$ ~ 200 – 250 kpc) from COS UV surveys [21, 34] finds another 10–20%. Therefore, at least 50% of the expected baryons associated with a $1 - 2 \times 10^{12} M_\odot$ DM halo are not accounted for. Their location and physical state is expected to be a very powerful diagnostic of physics involved in galaxy formation.

The mismatch of metals observed within galaxies with the predicted yields from stellar populations [35] indicates that a significant fraction — probably most — of metals reside in the CGM. Metals detected in UV quasar absorption spectra account for ~ 10% of the total output from star formation [36]. Similarly, the significant gaseous halo O VI masses observed by COS [24] are only a small percentage of the total circumgalactic oxygen budget. Cooling functions of hot gas predict far more O VII and O VIII [37], which will dominate the oxygen budget above $M_{\rm vir} = 10^{12} M_\odot$ (Fig. 1, lower panel). These ions are detectable at high abundance in our own halo though with poorly constrained distances [38]. Only by observing them around a variety of galaxies as a function of impact parameter can we reveal the amount and location of these invisible circumgalactic metal reservoirs.

The gaseous halo energy budget, which should be dominated by $T \geq 10^6$ K gas, is unconstrained by at least an order of magnitude. Completely absent from such an energy census are turbulent velocities [39], bulk motions, and rotation [40] of the hot gas, as well as magnetic fields [41] and non-thermal particles [42], which are frontiers for future observational facilities.

Knowing the mass, metals, and energy budget will depend on measurements of the detailed thermodynamic state of the hot halo — the dominant medium by volume and mass. Required observations include radial profiles of gas density, temperature, and metallicity. In addition, it is paramount that future facilities are able to *map* the CGM. The expected structure of all CGM components is very rich and directly related to dynamical processes in play (superwind outflows, acretion, etc., see Fig. 2). To catch this physics and to avoid incorrect inferences on the bulk properties of the CGM, individual objects must be observed and mapped to understand the asymmetries of the hot phase and how they relate to the properties of cool structures.

***Future observational probes of the CGM.*** UV absorption line studies of the CGM will continue as the complete COS legacy archive of absorbers will be cross-correlated with deep ground-based galaxy follow-up surveys. However, the next frontier will be observations of the hot phase. UV studies fundamentally cannot provide direct information on this phase because the gas is ionized above the UV transitions for relevant range of temperatures.

An upcoming new observational tool is sensitive Sunyaev-Zeldovich effect measurements. Detection of individual galactic halos in SZ is impossible even with future-generation experiments, but detections of the stacked thermal SZ effect have already been made with [43], and kinetic SZ effect may have been detected with Atacama Cosmology Telescope [ACT 44]. These detections are forecasted to become routine in the upcoming CMB "Stage 3" and "Stage 4" experiments [45], although the focus is on objects with high masses ~ $10^{13}$ $M_\odot$. Angular resolution also will be an issue for this work. The optimal redshift range for stacking SZ signal is at $z$ ~ 0.5 or above. At this redshift, a 1 arcmin beam of the "Stage 4" CMB experiments corresponds to 350 kpc, too coarse to constrain the structure of the CGM within the virial radius even in a stack.

**Therefore, the on-going and future studies of the CGM using the UV absorption lines and SZ effect are important, but the all-important detailed data for the hot phase of the CGM in individual objects must be obtained by another method. Future X-ray observatories can provide the measurements with required sensitivity, spectral, and spatial resolution.** These measurements will be augmented by observations of non-thermal phenomena (relativistic particles, magnetic fields, jets) with future radio astronomy facilities such as the Square Kilometer Array or ngVLA.

A reasonable minimum set of goals for future X-ray observations includes deep images in the soft X-ray bands for characterizarion of multiple modes of feedback operating inside $0.25R_{\rm vir}$ (50 kpc) of a $10^{12} M_\odot$ halo, and of structure of hot halos out to $0.5R_{\rm vir}$ (150 kpc) of a $10^{12.5} M_\odot$ galaxy. Detections in multiple bands over ~ 0.4–1.5 keV range will be needed to determine independently the CGM temperature, density, and metallicity. Spectroscopic observations are also essential. Sensitive absorption line measurements — at spectral resolving powers $R > 5000$ only accessible with X-ray gratings — can extend characterization of the CGM in MW-mass galaxies to at least the virial radius and provide additional kinematic data [46]. X-ray microcalorimeter measurements in the inner halo can resolve structures associated with feedback and gas accretion onto galaxies. The microcalorimeter should have at least an ≈ 0.3 eV energy resolution to observe gas motions with the expected of order ≲ 100 km/s velocities, and provide 1 arcsec spatial sampling to resolve expected structures [47].

The limiting factor for imaging observations of the CGM is its expected low contrast relative to the astrophysical and instrumental backgrounds. *Chandra*'s capabilities are insufficient by 1.5–2 orders of magnitude. For future missions, imaging of the CGM at $E \gtrsim 0.7$ keV will be severely affected unless most of the Cosmic X-ray Background (CXB) is resolved into discrete sources, for which arcsecond resolution is required. In addition, X-ray mirrors should be carefully baffled to block stray light or otherwise it will plague the measurements by introducing additional large-scale nonuniformities in the background. Unfortunately, stray light and residual CXB will fundamentally limit *Athena*'s ability to map diffuse gas in galactic halos and Cosmic Web filaments. *Athena* lacks X-ray gratings, and its microcalorimeter provides a resolving power of only $R$ ~ 300 at OVII, insufficient to measure the expected ~ 100 km/s gas velocities. *Lynx* mission design makes drastic improvements through a powerful package of outstanding imaging sensitivity, X-ray gratings with $R > 5000$, and a microcalotimeter subarray sporting 1″ pixels and 0.3 eV energy resolution. The needs of future CGM observations will be uniquely met by *Lynx*.

*Conclusion.* A sensitive X-ray telescope such as *Lynx* can bring about a revolution in observations of the CGM. Its hot component, now predominantly undetected, will be richly mapped around normal galaxies. These observations, augmented by advances in the radio and a continued progress in the UV and mm-waves, will expose the missing pieces of our understanding of galaxy formation.